# Optical cloaking with metamaterials

F. Bilotti, S. Tricarico, and L. Vegni

Department of Applied Electronics – University "Roma Tre"
Via della Vasca Navale, 84 – 00146 – Rome – ITALY
**e-mail**: vegni@uniroma3.it

**Summary**
In this contribution, we present the design of cylindrical electromagnetic cloaks working at optical frequencies, making use of layered structures of plasmonic and non-plasmonic materials. The simulated results confirm the validity of the proposed approach and show a rather broad-band behavior of the cloaking device.

**Introduction**
Different approaches to design electromagnetic cloaking devices have been proposed in the last few years by some groups worldwide [1]-[7]. Each of them is characterized by advantages and drawbacks. The approach proposed in [4]-[5] has some undoubted advantages in terms of the bandwidth of operation and the relatively simple nature of the material cover used to cloak a given object. In that case, in fact, the cover materials are *homogeneous* and characterized by a close-to-zero real part of the permittivity. This is an important advantage over, for instance, the method proposed in [1]-[3], which is based, instead, on the employment of *inhomogeneous* material cloaks with a certain profile of the constitutive parameters.

Even if the approach presented in [4]-[5] makes use of homogenous materials, the design of proper materials exhibiting the required values of permittivity is, anyway, a difficult task. In the microwave frequency range, a simple design of a cylindrical cloak has been recently proposed in [5]. The material with a close-to-zero real permittivity has been implemented in that case through a proper generalization of the parallel plate medium approach. Anyway, if we would like to extend the cloaking approach proposed in [4]-[5] to the visible regime, we run in troubles. On one hand, in fact, the parallel plate medium cannot be used anymore, and, on the other hand, there is a lack of useful natural materials. As previously anticipated, in fact, useful materials are characterized by a close-to-zero real permittivity, which means that the plasma frequency should rest in the visible. Unfortunately, noble metals, such as gold and silver, have plasma frequencies at smaller wavelengths in the ultra-violet regime and, thus, exhibit a strong plasmonic (i.e. the real permittivity is negative) behavior in the visible. For this reason, they are not useful as such.

In this contribution, we propose a way to extend the approach presented in [4]-[5] to the optical frequency range, through the employment of proper layered structures made of plasmonic and non-plasmonic materials. The same method can be straightforwardly applied also to obtain cloaking devices in the IR regime.

**Formulation and design**
In order to synthesize a material with the plasma frequency laying in the visible, we may use the approach proposed for a different purpose and for a planar configuration in [8]. By stacking two different material slabs, if the thicknesses of the slabs are electrically small, in [8] it is shown that the resulting composite material is described through constitutive parameters depending only on the ratio between the thicknesses of the two slabs and the constitutive parameters of the two different materials.

Now, let's suppose we want to cloak a cylindrical object with given permittivity and permeability ($\varepsilon_{obj}, \mu_{obj}$). Extending the formulation proposed in [8] to the case of the cylindrical geometry, we have the two configurations depicted in Fig. 1.

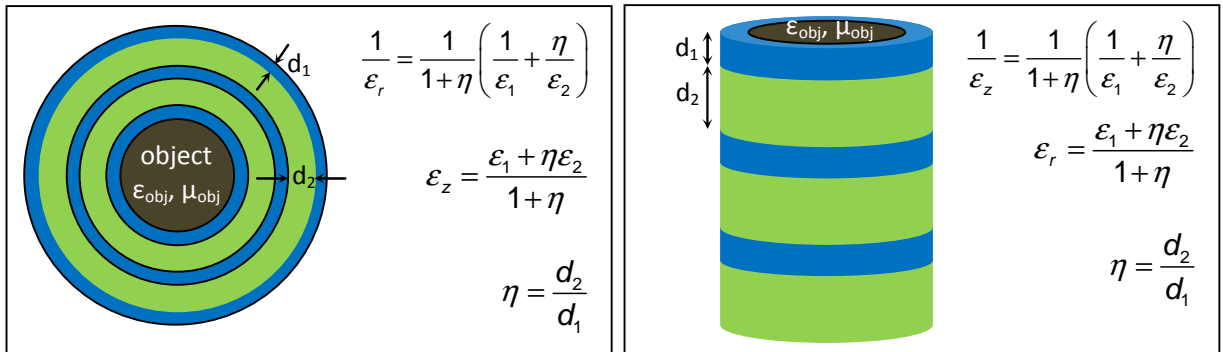

Fig. 1 *(left)* Cylindrical cloak made of concentric shells of different materials. *(right)* Cylindrical cloak made of stacked disks of different materials.

If the two materials are a plasmonic material like silver (Ag) and regular silica (SiO$_2$), it is possible to obtain in both configurations a close-to-zero real permittivity along the axis of the cylinder in the IR and visible regimes. Therefore, we can now apply the formulation in [4]-[5] to reduce the total scattering cross section of the covered cylindrical object in order to design the cloak.

In Fig. 2 we show the full-wave simulated results obtained in the case of a finite cylindrical object. The reduction of the total scattering cross-section is well evident.

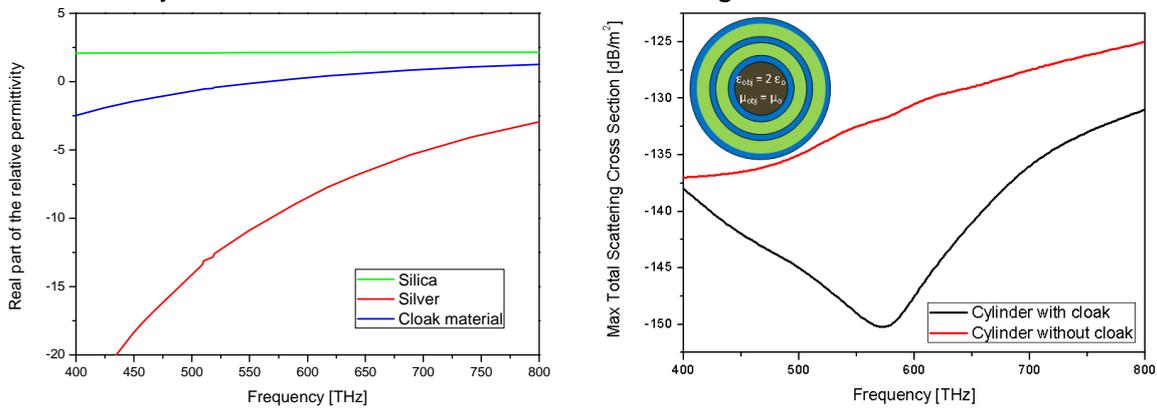

Fig. 2 *(left)* Real part of the permittivity of Ag, SiO$_2$ and the effective medium of the cloak along the axis of the cylinder. The cloak is designed for *f = 600 THz* and the thicknesses of the two material (Ag and SiO$_2$) shells are, *2 nm* and *10 nm*, respectively. *(right)* Maximum total scattering cross section of a cylindrical object (height = *300 nm*, radius = *50 nm*, $\varepsilon_{obj} = 2\varepsilon_0$, $\mu_{obj} = \mu_0$) with and without the cloak.